# New Astrometric Measurements of the Position Angle and Separation of the Double Star System WDS 03245+5938 STI 450

Miracle Chibuzor Marcel[1], Idris Abubakar Sani[2], Jorbedom Leelabari Gerald[3], Privatus Pius[4], Ohi Mary Ekwu [5], Bauleni Bvumbwe[6], Esaenwi Sudum[7], Joy Ugonma Olayiwola [8]

1 Pan-African Citizen Science e-Lab, FCT, Abuja: miracle.c.marcel@gmail.com

2 Centre for Basic Space Science and Astronomy-NASRDA: idris.abubakar@nasrdacbss.com

3 Pan-African Citizen Science e-Lab, FCT, Abuja; jorbedomlg94@gmail.com,

4 Department of Natural Sciences, Mbeya University of Science and Technology, Iyunga, Mbeya, 53119, Tanzania; privatuspius08@gmail.com,

5 Pan-African Citizen Science e-Lab, FCT, Abuja; ohipeaceao@gmail.com

6 Celestial Explorers, Blantyre, Malawi; baulenibvumbwe13@gmail.com

7 Rivers State University, Port Harcourt: esaenwi.sudum@ust.edu.ng

8 National Space Research and Development Agency, NASRDA HQ: jolayiwola@gmail.com

**Abstract**

We report new measurements of the position angle and separation of the double star WDS 03245+5938 STI 450, based on our observations, Gaia EDR3, and historical data. We find that the position angle and separation are 209.7° and 7.68", respectively, showing slight changes from the previous values of 210° and 7.742". We also find that the distances between the two stars are far apart, suggesting that the system is an optical double, and therefore not gravitationally bound together. Furthermore, we find that the ratio of proper motion (rPM) metric of the system is distinct, indicating that the system is a chance alignment of two unrelated stars that are at different distances.

## 1.0 Introduction

Double stars are a fascinating aspect of stellar astronomy, as they consist of two stars that appear to be close to each other when observed in the sky. These double stars can be classified into two categories: optical doubles and binary stars. Optical doubles are pairs of stars that are not physically related or gravitationally bound but only appear to be near each other along our line of sight. On the other hand, binary stars are pairs of stars that are gravitationally bound and orbit around their common center of mass (Chen X et al., 2023).

Binary stars are more common than single stars in our Milky Way galaxy (Kaib and Raymond, 2014). They are crucial in understanding stellar evolution, mass transfer, and gravitational interactions (Li and Han, 2008). By studying double stars, astronomers can gain valuable insights into the formation and evolution of stars.

This paper investigates the double-star system WDS 03245+5938 STI 450, using various sources of data. We analyzed historical data of the system from the Washington Double Star Catalog (Mason et al., 2001), as well as new data from the Las Cumbres Observatory and the Gaia EDR3 mission (Gaia Collaboration et al., 2022). The Las Cumbres Observatory provides high-quality observational data, while the Gaia mission offers precise astrometric measurements of celestial objects.

Our primary objective was to measure the position angle and separation of the two stars in the WDS 03245+5938 STI 450 system. These measurements allow us to determine the relative positions and separation between the stars. To ensure the accuracy of our results, we compared our findings with previous studies.

Furthermore, we aim to discuss the nature and classification of the WDS 03245+5938 STI 450 system. Based on our observations and analysis, we estimate whether it is an optical double or a binary star. This classification is crucial in understanding the dynamics and characteristics of the system.

### 1.1 Target Selection

To select the double-star system of interest, we used Stelle Doppie, a website that can access the database of the Washington Double Star Catalog (WDS) for double-star systems. We used a right ascension (RA) range of 17 hours and above and down to 7 hours so that the systems would be observable during the time of this study. We monitored this using Stellarium, a free software that simulates the sky. We had no restrictions on declination (Dec) because there are several 0.4m LCOGT networks mounted on both hemispheres. We selected stars that were observed before 2015, to ensure that the system's astrometry changed significantly. We chose the magnitude of the primary star to be between 9 and 11 so that it could be observed by the 0.4m LCO with a limit of 20.5 magnitudes. We did not specify the magnitude of the secondary star, because we wanted to have a difference in magnitude (Δmag) of less than 4 so that both stars would be visible. We chose the separation to be between 5 and 10 arcseconds so that they could be easily resolved as two separate stars in images taken by the LCO. We had 223 search results from which we selected STI 450 as a worthy double-star system.

The STI 450 double star system was chosen for several reasons:

1. Since the first observation in 1903 and the last observation in 2003, they have only been studied four times.

2. The nature of this system remains uncertain.

3. It has been twenty years since the most recent measurement of the star, providing some time for the two stars to have moved relative to each other.

### 1.2 Background

WDS 03245+5938 STI 450 is a double star system in the constellation Camelopardalis, with the equatorial coordinates of RA = 03h 24m 29.25s and DEC = +59° 38' 41.8". Figure 1 shows its position in the sky.

STI 450 was first identified as a double star in 1903. Since then, the system has been studied four times, most recently by Hartkopf et al. (2013). The system was also included in two large-scale surveys of stellar astrometry and photometry: the AC 2000 catalog (Urban et al., 1998), which contains the positions and magnitudes of more than 4.6 million stars, and the USNO CCD Astrographic Catalog (UCAC4) (Hartkopf et al., 2013), which provides high-precision positions and proper motions for more than 113 million stars. Both catalogs can be used to search for and measure double stars. According to the Washington Double Star Catalog (WDS), the primary component of STI 450 has an apparent magnitude of 9.87 and a spectral type of F5, indicating that it is a white main-sequence star with a surface temperature of about 6,000 to 7,500 K and a luminosity of about 1 to 5 times that of the Sun. The secondary component has an apparent magnitude of 13.60 and a spectral type of G0, indicating that it is a yellow main-sequence star with a surface temperature of about 5,500 to 6,000 K and a luminosity of about 0.8 to 1.2 times that of the Sun. The latest measurement of the system as indicated at the WDS website reported a position angle of 210° and a separation of 7.742 arcseconds. However, given the time elapsed since the last observation (2003 – 2023), we aim to provide a new solution for the position angle and separation of the system.

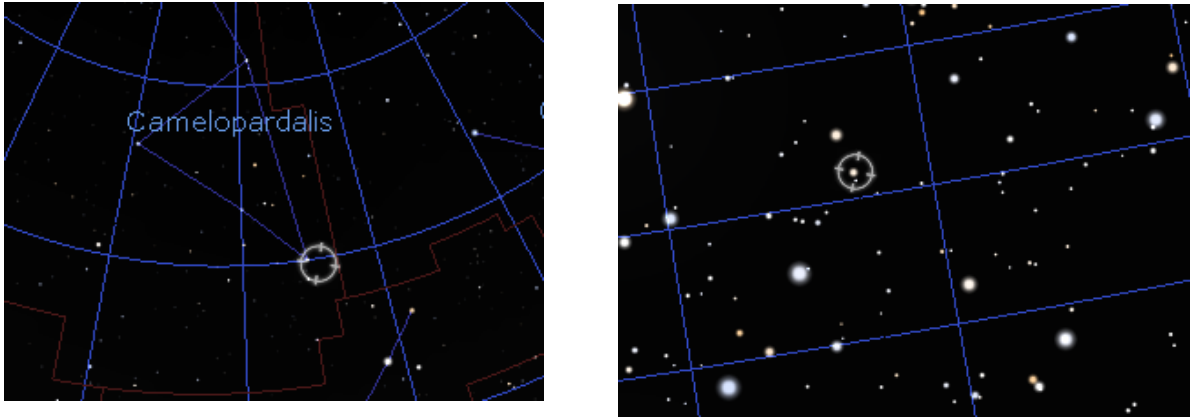

*Figure 1. Location of WDS 03245+5938 STI 450 in the constellation Camelopardalis (left); image of the surrounding starfield, as shown in Stellarium (right).*

**2.0 Methods and Observations**

On October 11, 2023, we used the 0.4m Haleakala Observatory, which is part of the Las Cumbres Observatory's Global Telescope (LCOGT) network, to take 10 images of our system with an exposure time of 2 seconds each. The observatory has SBIG 6303 cameras, which have a pixel scale of 0.571 arcsec pixel and a field of view of 29′x19′. We used the Bessell-V filter for our observations. The LCO processed all image files automatically using their BANZAI pipeline. We then used the AstroImageJ program (Collins, 2017) to measure the separation and position angle of our system from the images. We did this by choosing an aperture size of 5 pixels. AstroImageJ allows the user to find the centroid of the star automatically using the position that represents the weighted average of the pixel brightnesses within a chosen aperture. Our aperture size is large enough to enclose the star, but not so large that the apertures of the two stars overlap. After selecting the aperture size and zooming into the region of the image where the stars are located, the user command-drags the cursor from the primary to the secondary star to obtain the PA and Sep (shown as ArcLen). The LCOGT and our image measurements are shown in Figures 2 and 3 respectively.

We also requested historical data for the systems from Dr. Rachel Matson at the Washington Double Star Catalog and plotted the new and historical measurements together using Google Sheets.

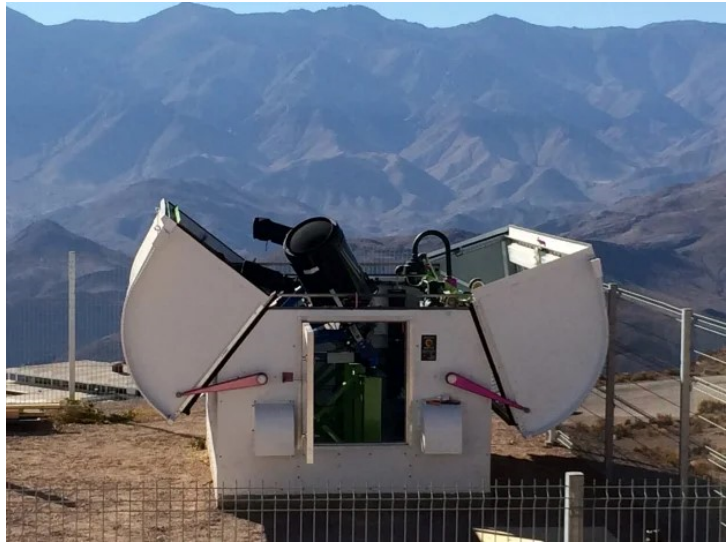

*Figure 2: 0.4 m telescope located at one of Las Cumbres Observatory's sites.*

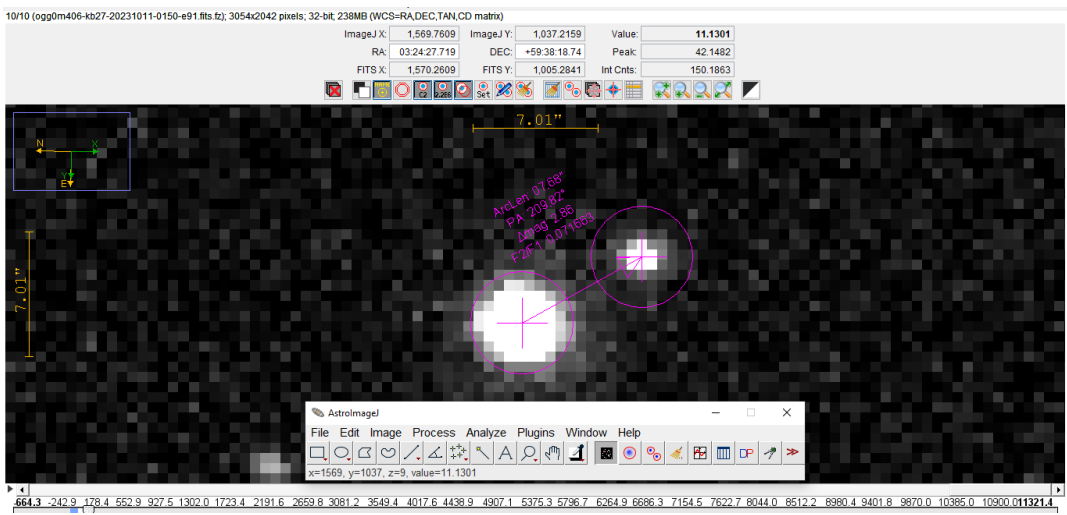

*Figure 3: Sample image of WDS 03245+5938 STI 450 double star system from AIJ.*

### 3.0 Data

Table 1 shows the new measurements derived from our ten images. Table 2 is a summary of statistics for our measurements. Table 3 shows Gaia data for our target double-star system, and Table 4 lists the systems' historical data sent by the Washington Double Star Catalog.

Figure 4 shows the plot of WDS historical data and the new measurements of the double star system.

Table 1: New measurements of WDS 03245+5938 STI 450 derived from our images

| S/N | PA (°) | Sep (") |
|---|---|---|
| 1 | 209.4 | 7.71 |
| 2 | 210.1 | 7.61 |
| 3 | 209.6 | 7.68 |
| 4 | 209.3 | 7.57 |
| 5 | 209.3 | 7.77 |
| 6 | 209.6 | 7.74 |
| 7 | 210.3 | 7.79 |
| 8 | 209.9 | 7.69 |
| 9 | 209.5 | 7.60 |
| 10 | 209.8 | 7.68 |

Link to our images

Table 2: Mean, standard deviation, and Standard error of the mean of our measurements WDS 03245+5938 STI 450

| Double Star | Date | Images | | PA (°) | Sep (") |
|---|---|---|---|---|---|
| WDS 03245+5938 STI 450 | 11th of October, 2023 (2023.7781) | 10 | Mean | 209.7 | 7.68 |
| | | | Standard Deviation | 0.34 | 0.068 |
| | | | Standard Error of the Mean | 0.11 | 0.022 |

Using data from the Gaia Data Release 3 (DR3), we retrieved the parallax values and proper motion of each star and calculated their rPM metric, which is used to compare their proper motions. These data are shown in Table 3.

Table 3: Parallax and Proper Motion data from the Gaia DR3 and the rPM metric

| Parallax of Primary star (mas) | Parallax of Secondary star (mas) | Proper Motion of Primary (mas/yr) | Proper Motion of Secondary (mas/yr) | rPM |
|---|---|---|---|---|
| 1.707 | 5.5932 | pmra = - 6.262 | pmra = - 6.960 | 0.866 |
| Distance of Primary star in parsecs | Distance of Secondary star in parsecs | pmdec = - 0.400 | pmdec = 7.627 | |
| 585.8 | 178.8 | | | |

The proper motion (PM) of a star is shown in columns 3 and 4 of Table 3, in RA and Dec. These numbers were used to calculate the ratio of proper motions (rPM) metric. The rPM, which was an idea by Harshaw (2016) (equation below), is the magnitude of the difference between the proper motions of the primary and secondary divided by the larger magnitude of the two proper motions. In other words, the rPM measures how different the two proper motions are from each other, relative to the size of the larger proper motion. If the rPM of the stars is less than 0.2, then the stars are likely a Common Proper Motion (CPM) pair. If the rPM is less than 0.6, then the stars have Similar Proper Motion (SPM), and if the rPM is greater than 0.6, then they have Distinct Proper Motion (DPM) (meaning different or non-common). Going by this analogy, STI 450, having an rPM metric value of 0.866, shows that the system has Distinct Proper Motion (DPM) pairs.

$$\text{Resultant} = \sqrt{(R_{pri} - R_{sec})^2 + (D_{pri} - D_{sec})^2}$$

$$\text{Vector} = \sqrt{R^2 + D^2}$$

$$\text{rPM} = \frac{Resultant}{Vector}$$

Where $R_{pri}$ & $D_{pri}$ represent the Right Ascension and Declination of the primary star, $R_{sec}$ & $D_{sec}$ represent the Right Ascension and Declination of the secondary star, R & D represent the Right Ascension and Declination of either the primary and secondary star, but whose vector is higher.

Using the WDS historical data of the system as shown in Table 4, we made a plot of the system. This plot, presented in Fig 4, displays the data points in the color pattern ROYGB, where the right ascension is on the x-axis and the declination is on the y-axis. The first observation data point of the system made in 1903 is represented in red (R), and our new measurement in 2023 is represented in blue (B).

Table 4: WDS 03245+5938 STI 450 historical data from the Washington Double Star Catalog. They have different numbers of decimal places. link to data

| Year | PA (°) | Sep (") |
|---|---|---|
| 1903.94 | 221.1 | 7.5 |
| 1911.02 | 209.1 | 7.9 |
| 1998.98 | 209.4 | 7.61 |
| 2003.697 | 210.0 | 7.742 |
| 2023.7781 | 209.7 | 7.68 |

Figure 4 shows the plot of historical data and the new measurements of the double star system. Red, Orange, Yellow, and Green are the historical data points. Blue is the data point from our measurements.

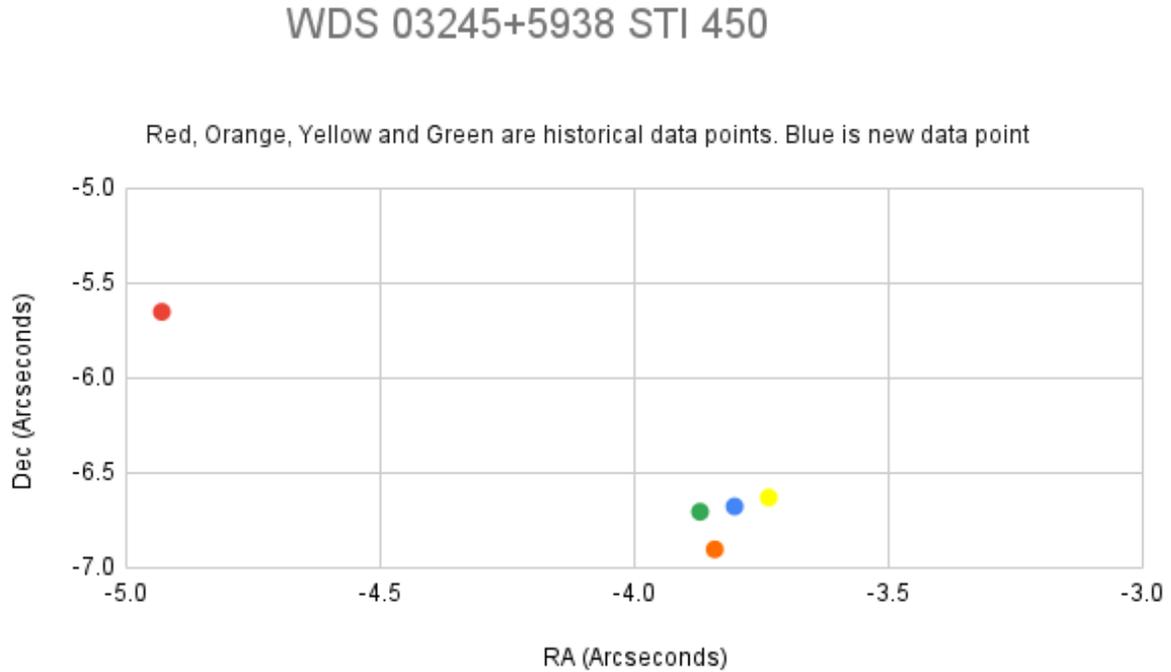

*Figure 4: Plot of historical data and the new measurements of WDS 03245+5938 STI 450*

Link to table

**4.0 Discussions**

We present the results of our study of WDS 03245+5938 STI 450, a double star system, with the new position angle and separation as 209.7° and 7.68" respectively. Based on historical data, the Gaia Data, and our new measurements, we show the variation in the separation of the two stars in Table 4 and plotted in Figure 4

Table 4 when traced from 1903 to 2023, shows that the separation between the two stars increases and decreases in a non-uniform pattern, also seen from Figure 4, according to the color pattern ROYGB. this is largely attributable to atmospheric effects, noise in the data, and general measurement uncertainty. Therefore, Our study of STI 450 is inconclusive both in terms of the new measurements and the pre-existing data

Another striking feature of the system consisting of the primary and secondary stars is the large discrepancy in their distances from Earth, as measured by their parallax values. The secondary star is more than three times farther away than the primary star, with parallax values of 5.5932 mas and 1.707 mas, respectively, or distances of 178.8 parsecs and 585.8 parsecs. This indicates that the stars are far apart and could possibly not be gravitationally bound. This leads us to conclude that the system is an optical double, rather than a binary system.

To support our point, according to the rPM metric of the system in Table 3, we derived from the proper motions (in RA and Dec) of the double star system, which gave the value $0.866$, showing that the system has Distinct Proper Motion (DPM). This non-commonality suggests that the system is very likely to be a chance alignment of two unrelated stars that are at different distances

## 5.0 Conclusions

In this study, we have presented updated measurements of the position angle and separation of the double star WDS 03245+5938 STI 450, based on our observations, the Gaia Data, and historical data. We have shown that the separation of the two stars varies in a non-uniform pattern over time, which could be due to atmospheric effects, noise in the data, and general measurement uncertainty. Therefore, we cannot ascertain the nature of the system based on the historical records and our measurements.

The data of the system retrieved from the Gaia EDR3 catalog showed that the parallax values of the two stars are significantly different, indicating that the system is an optical double, rather than a binary system, and consequently not gravitationally bound.

Moreover, we have found that the rPM metric of the system is distinct also indicating that the system is just a chance alignment of two unrelated stars that are at different distances in space.

**Acknowledgments**

This research was made possible by the Washington Double Star catalog maintained by the U.S. Naval Observatory, the StelleDoppie catalog maintained by Gianluca Sordiglioni, Astrometry.net, and AstroImageJ software, which was written by Karen Collins and John Kielkopf.

This work has also made use of data from the European Space Agency (ESA) mission Gaia (https://www.cosmos.esa.int/gaia), processed by the Gaia Data Processing and Analysis Consortium (DPAC, https://www.cosmos.esa.int/web/gaia/dpac/consortium). Funding for the DPAC has been provided by national institutions, in particular, the institutions participating in the Gaia Multilateral Agreement.

This work makes use of observations taken by the 0.4m telescopes of the Las Cumbres Observatory Global Telescope Network located at Mt. Haleakala, Hawaii, USA.

We, the Pan-African Citizen Science e-Lab (PACS e-Lab) astronomy research and publication group, whose aim is to spread astronomy in Africa through engagement in citizen science projects and astronomy research, would also like to thank Dr. Rachel Freed of the Institute for Student Astronomy Research (InStAR) for her consistent support, time, and guidance throughout this project, Gianluca Sordiglioni for running the wonderful site StelleDoppie, where we obtained our base information about our star systems, and lastly, Kalee Tock for creating the plotting instructions.